\documentclass[preprint,aps,floats,subeqn,showpacs,amssymb]{revtex4}
\usepackage{epsfig}

\newcommand{\be}{\begin{equation}}
\newcommand{\ee}{\end{equation}}
\newcommand{\bea}{\begin{eqnarray}}
\newcommand{\eea}{\end{eqnarray}}
\newcommand{\bml}{\begin{mathletters}}
\newcommand{\eml}{\end{mathletters}}

\begin{document}

\tighten





\title{Black holes, black strings and cosmological constant} 
\author{Y. Brihaye \footnote{yves.brihaye@umh.ac.be}
}  
\affiliation{Facult\'e des Sciences, Universit\'e de Mons-Hainaut, 7000 Mons, Belgium}
\date{\today}
\setlength{\footnotesep}{0.5\footnotesep}
\begin{abstract}
We present a review of black holes and black string solutions available in the $d$-dimensional
Einstein and Einstein-Maxwell model in the presence of a cosmological constant.  
Due to the cosmological constant,  the equations do not admit explicit
solutions for generic values of the parameters and numerical methods are necessary to
construct the solutions.   Several new features of the solutions are discussed,
namely their stability and the occurrence of non-uniform black strings which depend non-trivially
on the co-dimension.
Black string solutions are further constructed for the Einstein-Gauss-Bonnet model. 
The influence of the Gauss-Bonnet term on the domain of existence of the black strings is  discussed in details.  \\
\\
{\it Extended version of a contribution to the 418th WE-Heraeus Seminar, Bremen, 25-29 August 2008.}
\end{abstract}
\maketitle 
\section{Introduction}
In the last years, there has been increasing 
interest in space-times involving more than four dimensions. Particularly,
so called brane-world models \cite{brane,rs} have gained a lot of
interest. These assume the standard model fields to be
confined on a 3-brane embedded in a higher dimensional manifold.
Consequently, a large number of higher dimensional black holes has been studied in 
recent years. The first solutions that have been constructed
are the hyperspherical generalisations of well-known black holes
solutions such as the Schwarzschild and Reissner-Nordstr\"om solutions
in more than four dimensions \cite{tangherlini} as well
as the higher dimensional Kerr solutions \cite{mp}.
In $d$ dimensions, these solutions have horizon topology $S^{d-2}$.

In contrast to four dimensions, however, black holes with different horizon
topologies should be possible in higher dimensions. 
An example is a 4-dimensional
Schwarzschild black hole extended into one extra dimension, a so-called
Schwarzschild black string. These solutions have been discussed extensively
especially with emphasis put on their stability \cite{gl}.
A second example, which is important due to its implications for
uniqueness conjectures for black holes in higher dimensions is the black ring
solution in 5 dimensions with horizon topology $S^2\times S^1$ \cite{er}.

On the other hand there is mounting observational evidence in the past few years
\cite{astrocc} that the universe is  expanding with acceleration. 
The simplest explanation for this is  a positive cosmological constant.
From  a more theoretical point of view, the Anti--de--Sitter/Conformal Field Theory (AdS/CFT) correspondence 
 \cite{Witten:1998qj,Maldacena:1997re} encourages
the investigations of the field equations in the presence of a negative 
cosmological constant. It therefore make sense to investigate the effects
of a cosmological constant -- either positive or negative -- on black objects. 
The results of this investigation are discussed in this report.

In the first part (Section II) of this report, we discuss rotating black hole solutions
of the $d$ dimensional Einstein-Maxwell model (with $d$ odd). The ansatz chosen
for the metric and Maxwell fields leads to a set of differential equations. The
domain of existence of these solutions is determined in dependence on the horizon
radius and on the strength of the magnetic field. Section III is devoted to several 
aspects of black strings. In the presence of a negative cosmological constant, these
solutions can not be given in explicit form, but have to be constructed numerically.
In particular, charged and rotating black strings are considered. The stability
of AdS black strings is then discussed and it is shown that they become unstable when
the length of the extra dimension gets larger than a horizon-dependent critical value.
Preliminary results suggesting the existence of non-uniform black strings, depending both
on the extra dimension and on the radial variable associated to the internal space-time,
are presented.
In the last section  various properties of black string solutions
of the Einstein-Gauss-Bonnet model are presented. The influence of the
Gauss-Bonnet interaction on the domain of existence of the black strings is analyzed
in detail.

\section{Black holes with cosmological constant}
\label{sec:1b}

In this section we consider the Einstein-Maxwell equations in $d$ dimensions and
address the construction of charged, rotating black holes. For odd values of $d$,
an ansatz can be done which transforms the full equations into a system of ordinary differential
equations. Numerical investigation of the solutions is then possible.
 The
domain of existence of these solutions is determined depending  on the horizon
radius and on the strength of the magnetic field. 
\subsection{Model and equations}
The Einstein-Maxwell Lagrangian  with a cosmological 
constant $\Lambda$ in a $d-$dimensional space-time is given by 
\begin{eqnarray}
\label{action-grav}
I=\frac{1}{16 \pi G_d}\int_M~d^dx \sqrt{-g} (R - 2\Lambda- F_{\mu \nu}F^{\mu \nu}) \nonumber \\
-\frac{1}{8\pi G_d}\int_{\partial M} d^{d-1}x\sqrt{-h}K,   
\end{eqnarray}
Here, $G_d$ denotes the $d-$dimensional Newton constant. The units are chosen in such a way
that $G_d$ appears as an overall factor. The last term in (\ref{action-grav}) is the Hawking-Gibbons term;
it guarantees the variational principle to be well defined but we will not need it in this report.
It is convenient to further define an (Anti-)de-Sitter ``radius''  $\ell$ according to 
\be
\Lambda = \pm \frac{(d-2)(d-1)}{2 \ell^2}. 
\ee
The Einstein-Maxwell equations are obtained from the variation of the action 
with respect to the metric and the electromagnetic fields.

\textbf{{The ansatz}:} We consider space-times with odd dimensions, $d=2N+1$, and assume the
metric to be of the form
\begin{eqnarray}
\label{metric}
 ds^2 = &-&b(r)dt^2 +  \frac{ dr^2}{f(r)} + 
g(r)\sum_{i=1}^{N-1}
  \left(\prod_{j=0}^{i-1} \cos^2\theta_j \right) d\theta_i^2  \\
 &+& h(r) \sum_{k=1}^N \left( \prod_{l=0}^{k-1} \cos^2 \theta_l
  \right) \sin^2\theta_k \left( d\varphi_k - w(r)
  dt\right)^2 \nonumber  \\
 &+&  p(r) \left\{ \sum_{k=1}^N \left( \prod_{l=0}^{k-1} \cos^2
  \theta_l \right) \sin^2\theta_k  d\varphi_k^2 \right.
  -\left. \left[\sum_{k=1}^N \left( \prod_{l=0}^{k-1} \cos^2
  \theta_l \right) \sin^2\theta_k   d\varphi_k\right]^2 \right\} \ \nonumber.
\end{eqnarray}
This metric possesses $N+1$ Killing vectors $\partial_{\varphi_k}$,$\partial_t$, out of which
$N$ are associated to conserved angular momenta.
The most general Maxwell potential consistent with these symmetries turns out to be
\begin{equation}
\label{Maxwell}
 A_\mu dx^\mu=V(r)dt+a_\varphi(r) 
\sum_{k=1}^N
\left( \prod_{l=0}^{k-1} \cos^2\theta_l \right) 
\sin^2\theta_k d\varphi_k          \nonumber
\end{equation}

The form of the metric (\ref{metric}) looks cumbersome, however, 
for $d=5$, it simplifies to 
\begin{eqnarray}
 ds^2 &=& -b(r)dt^2 +  \frac{ dr^2}{f(r)} + g(r) d \theta^2 +
  p(r) (\sin \theta)^2  (\cos \theta)^2    (     d\varphi_1 -  d\varphi_2)^2\ \\
&+& h(r) [(\sin \theta)^2 (d\varphi_1 -w(r) dt)^2 + (\cos \theta)^2  (d\varphi_2 -w(r) dt)^2] \nonumber \\
 \nonumber
\end{eqnarray}
 Inserting the ansatz above into the Einstein-Maxwell equations  results  in a system of seven
non-linear, coupled  differential equations provided $p(r) = g(r)-h(r)$.
The Maxwell functions $V(r), a_{\phi}(r)$ as well as the metric functions $b(r),f(r),g(r),h(r),w(r)$ are unknown.
One of these functions can be fixed arbitrarily, e.g. by choosing  the radial coordinate to be of the Schwarzschild type e.g. $g = r^2$.

\subsection{Explicit Solutions}
The set of differential equations under consideration admits a few explicit solutions 
in some specific limits. We will remind them before discussing the full solutions.\\
(i) {The vacuum black holes} 
are recovered for  vanishing Maxwell fields~: $V=a_{\varphi}=0$. The metric fields then take the form
\begin{eqnarray}
\label{vacuum}
 f(r)=1
-\epsilon \frac{r^2}{\ell^2}
-\frac{2M\Xi}{r^{d-3}}
+\frac{2Ma^2}{r^{d-1}},~ 
h(r)=r^2(1+\frac{2Ma^2}{r^{d-1}})   \ \  ,~ \nonumber  \\
w(r)=\frac{2Ma}{r^{d-3}h(r)},~~
g(r)=r^2,~~ b(r)=\frac{r^2f(r)}{h(r)} \ ,
\end{eqnarray}
where $M$ and $a$ are two constants related to the solution's mass and 
angular momentum and $\Xi \equiv 1+a^2/\ell^2$. 
These solutions generalize the Tangherlini \cite{tangherlini} and Myers-Perry \cite{mp} solutions 
to the case of non-vanishing cosmological constant. From now on, we will denote the 
angular velocity at the event horizon by $\Omega \equiv w(r_h)$.  

(ii) {The (Anti-)de-Sitter-Reissner-Nordstr\"om black holes} \cite{Gibbons:2004js} are recovered in the limit $w(r)=a_\varphi(r)=0$~:
\begin{eqnarray*}
f(r)&=&b(r)= 1 - \epsilon \frac{r^2}{\ell^2} - \frac{2 M}{r^{d-3}} + \frac{q^2}{2(d-2)(d-3) r^{2(d-3)}}  \ \ \ , \nonumber \\
 \ \ h(r)&=&g(r)=r^2 \ \ , \ \ 
V(r) = \frac{q}{(d-3)r^{d-3}} 
\end{eqnarray*}
where $M$ and $q$ are related to the mass and electric charge of the solution.

(iii) 
Charged, rotating black holes are  also known explicitely when a Chern-Simons term 
with a specific coupling constant is added to the model. These supersymmetric solutions are constructed
in \cite{ks} and \cite{clp}. In this review, we put the emphasis on the Einstein-Maxwell action 
with a minimal coupling of  gravity. 
We discuss first the solutions in the case of a positive cosmological constant.
\\
\subsection{Charged Rotating black holes for $\Lambda > 0$} 
 In this case, we expect a cosmological horizon to appear. The solution is therefore
  plagued with two  horizons at $r=r_h$ and $r=r_c$. In other words $f(r_h)=0$, $f(r_c)=0$, $b(r_h)=0$, $b(r_c)=0$.
 The equations therefore have two singular points and a strategy has to be implemented to deal
 with the numerical construction \cite{bd_2007}. It is elaborated along the following lines~:
 \begin{itemize}
 \item[.] Use a Schwarzschild coordinate {$g(r)=r^2$}. 
 \item[.] Fix $r_h$, $r_c$ by hand and add an equation  {$d \Lambda / dr = 0$}.
 \item[.] Implement the boundary conditions at $r_h$, $r_c$ and solve the equations for {$r \in [r_h,r_c]$}, determining $\Lambda$.
 \item[.] Solve Eqs. for {$r \in [r_c, \infty]$} as a Cauchy problem with initial data at $r=r_c$.
 \end{itemize}
\begin{figure}
\center
\includegraphics[height=6cm]{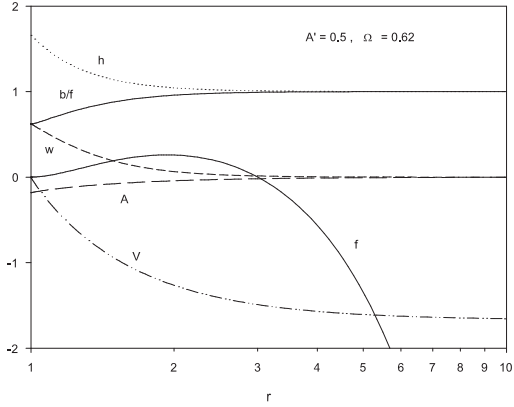}
%
%
\caption{The profile of the metric and Maxwell functions for $r_c=3,r_h=1$
for $a_h'=0.5$ and $\Omega=0.62$}
\label{f1}       
\end{figure}
 
The equations are cumbersome and it is not necessary to write them explicitely (see e.g. \cite{bd_2007}), 
for the purpose of this report,
we just write their overall structure which turns out to be~:
$$  \Lambda ' = 0 \ \ , \ \ f' = \dots \ , \ \ , b'' = \dots \ \ , \ \ h'' = \dots \ \ , \ \ 
  w'' = \dots \ \ , \ \ V'' = \dots \ \ , \ \ a_{\varphi}'' = \dots $$
  where the dots symbolize functions of $f,b,h,w,a_{\phi}$ and of the derivative $b',h',w',V',a_{\phi}$.  
The fields  $b,w,V$ can be arbitrarily rescaled according to  
\begin{equation} 
b \to \mu^2b \ \  , \ \ w \to \mu w \ \ , \ \  V \to \mu V + C \ \ , \ \ \mu, C \ {\rm constants.}
\end{equation}
After inspection of the equations and using this invariance, the boundary conditions for $r\in [r_h,r_c]$
can be choosen according to 
\begin{equation}
  f(r_h) = 0 \ \ ,  \ b(r_h) = 0 \ \ , \ b'(r_h) = 1  \ \ , \  \Gamma_h(r_h) = 0 \ \ 
     \end{equation}
\begin{equation}
 \ \  w(r_h) = w_h \ , \ \ V(r_h) = 0 \  , \ \ a'_{\varphi}(r_h)=a_h \  , \Gamma_A(r_h)=0 
\end{equation}
fixing  the arbitrary scale of $b$.
The parameters $w_h$, $a_h$ are fixed by hand and control the angular and magnetic moments respectively.
 At the cosmological horizon we set~:
\begin{equation}
  f(r_c) = 0 \ \ , \ \ b(r_c) = 0 \ \ , \ \ \Gamma_h(r_c) = 0 \ \ ,
  \ \ \Gamma_A(r_c) = 0 
\end{equation}
completing the set of fourteen conditions. 
The conditions $\Gamma = 0$ appearing in several  boundary conditions are necessary conditions
for the solutions to be regular at the  horizon. For instance, we find
\begin{equation}
\Gamma_A (r) \equiv 4 a_{\varphi} b' h + r^4 f'(h w' V' + a_{\varphi}' h w w' 
- a_{\varphi}' b')(r)
\end{equation}
and an even more involved expression for $\Gamma_h$.

\begin{figure}
\center
\includegraphics[height=6cm]{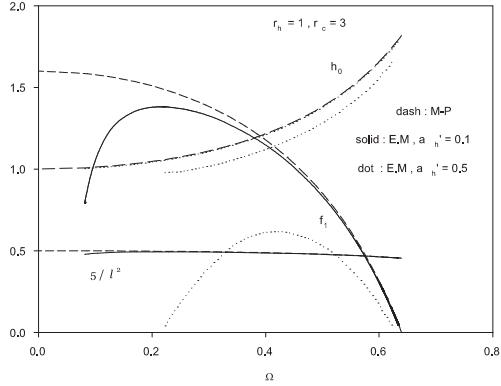}
%
%
\caption{Some metric parameters and the parameter $\ell^2$ as functions of $\Omega$
for   $r_c=3,r_h=1$ and for several values of  $a_h'$}
\label{f2}       
\end{figure}

\begin{figure}
\center
\includegraphics[height=6cm]{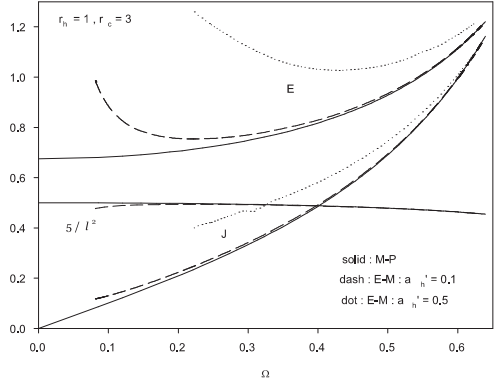}
%
%
\caption{Mass and Angular momentum of the $\Lambda > 0$ Black holes for $r_c=3$, $r_h=1$
as function of the angular velocity $\Omega$}
\label{f-2}       
\end{figure}

It should be stressed that 
 the functions $w,b,V$ have to be renormalized after the integration on $[r_h,\infty]$
 in such a way that space-time is asymptotically de Sitter. In particular
$$
     b(r) =  - \Lambda r^2 + 1 + O(1/r^2)  \ \ \ \ {\rm for}  \ \ r \to \infty
$$
As a consequence, it turns out to be impossible (at least it is extremely lengthy) to study the solutions for fixed charge $Q$
and varying $\Omega$.
The best way to study the domain of existence of solutions in the $a_h$--$\Omega_h$ plane  consists in fixing
the constant $a_h$ and vary the parameter $w_h$. After the suitable renormalisation of $b(r),w(r),V(r)$, 
families of rotating solutions with $a_h$ fixed can finally be constructed. A typical solution
is shown in Fig.1.
The numerical calculations further indicate the
following features~:\\
(i) For fixed $a_h$ and  varying $w_h$, black holes exist only on a finite interval of the
horizon angular velocity $\Omega$. \\
(ii) In the critical limits, 
solutions converge to extremal black holes, i.e. with $f(r_h)=0$, $f'(r_h)=0$.
These statements are illustrated by Fig. \ref{f2} .
\\
Let us remark that the value $1/\ell^2$ turns out to depend only weakly on $\Omega$;
it can therefore be considered that the family of rotating solutions
constructed with fixed $r_h,r_c$, $a_h$ correspond to  $\Lambda$ nearly constant.
The numerical results further indicate that, while approaching the  
boundary of the domain of existence of the black holes, the event horizon $r_h$ becomes extremal. 
This makes the numerical analysis difficult. However, extremal solutions can be constructed
directly by implementing the following trick: (i) we introduce an arbitrary scale, say $\alpha$ for the Maxwell field $a_{\varphi}$
and supplement the system with an equation $d\alpha / dr =0$, (ii) we take advantage of the extra equation
to replace the boundary conditions $f(r_h)=0$, $a'_{\phi}(r_h)=a_h$  by $f(r_h)=0 , df/dr(r_h)=0$, $db/dr(r_h)=0$. 
This produces extremal solutions with a definite value of $\alpha$.    
\\
{\textbf{Physical quantities:}}
Asymptotic global charges can be associated to each Killing vector of the metric. Using standard
results \cite{Balasubramanian:1999re,Brown:1993,ghezelbash_mann}, 
the mass-energy $E$  and the angular momentum $J$ of the solutions can be computed.
These physical quantities depend crucially on the asymptotics of the solutions. After some computation
we obtain
\begin{equation}
\label{asym}
b(r)=- \epsilon \frac{r^2}{\ell^2}+1+ \frac{\alpha}{r^{d-3}} +O(1/r^{2d-6}) \ \ , \ \ 
f(r)=- \epsilon \frac{r^2}{\ell^2}+1+\frac{\beta}{r^{d-3}} +O(1/r^{d-1}),
\end{equation}
\begin{equation}
h(r)=  r^2(1+  \epsilon \frac{\ell^2(\beta-\alpha)}{r^{d-1}} +O(1/r^{2d-4})) \ \ , \ \ 
w(r)=    \frac{\hat J}{r^{d-1}} +O(1/r^{2d-4 }). 
\end{equation}
where $\epsilon = \pm$ denotes the sign of the cosmological constant. 
From these expansions, the mass and angular momentum can be computed~: 
\begin{equation}
\label{grav-charges}
 E=\frac{V_{d-2}}{16\pi G_d}(\beta-(d-1) \alpha),
~~J=\frac{V_{d-2}}{8\pi G_d}\hat J \ .
\end{equation}
The dependence of $E,J$ on the angular velocity at the horizon $\Omega$ is illustrated
in Fig. \ref{f-2}.
Conserved quantities can be defined at the cosmological horizons as well.  
Smarr formulae relating them have been obtained \cite{bd_2008}.
\\
\subsection{Charged, rotating black holes with  {$\Lambda \leq 0$}}
\begin{figure}
\center
\includegraphics[height=6cm]{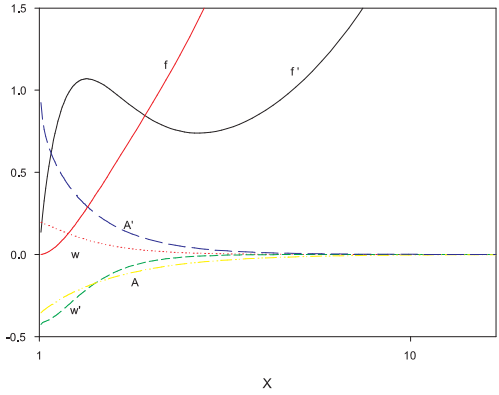}
%
%
\caption{Extremal rotating black hole with $\Omega = 0.2$ and $\Lambda < 0$}
\label{fig_extremal}       
\end{figure}

Solutions  with $\Lambda = 0$ are constructed numerically  in \cite{Kunz:2006eh} and in \cite{knlr}
where the isotropic coordinate is used toparameterizee the metric.
Charged, rotating black holes with  $\Lambda < 0$ are constructed in \cite{knlr} also
 using the isotropic coordinate.
We reconsidered several of these solutions using Schwarzschild coordinates.
  It is worth
stressing, however, that the patterns of solutions look different when solving the
equations  in the isotropic coordinate, say $y$ with $y_h$ fixed and 
in Schwarzschild coordinates $r$, with $r_h$ fixed, respectively.
 The pattern obtained for the case $\Lambda < 0$ (i.e. with $r_h$ fixed and
$\Lambda$ fixed) is very similar to the one suggested by Fig \ref{f-2}. In particular,
the solutions terminate into an extremal black hole (the event horizon is extremal) 
when a maximal value of $\Omega$ is reached. The profiles of such an extremal black hole
are presented in Fig. \ref{fig_extremal}.
\\
\section{Black strings with cosmological constant} 
\subsection{General setting}
In this section, we discuss black string solutions of the vacuum Einstein equations in $d$-dimensions and in the presence of  a cosmological constant. 
For this type of black objects, one of the spacelike dimensions
of the space-time  manifold, say {$z \equiv x_{d-1}$}, plays a special role~: space-time is chosen, 
a priori, as a warped product of a $d-1$-dimensional black hole metric with the extra-dimension $z$
which is assumed to be periodic with a period $L$. 
The horizon of the black string then has a topology of $S_{d-3} \times S_1$. 
The simplest case consists in assuming the metric to be independent of $z$, 
the corresponding solutions are then called uniform black strings (UBS).
The metric has the form
\begin{eqnarray}
\label{metricds} 
ds_{bs}^2=a(r)dz^2 + ds^2
\end{eqnarray}
where $ds^2$  is given by (\ref{metric}) (see previous section).
\\ In the case  $\Lambda = 0$  non-uniform black strings  are known to be unstable
\cite{gl} for  sufficiently large values
of $L$. Stable solutions can further beconstructedd with a  metric depending on $r$ and $z$.
They are called non-uniform black strings \cite{wiseman}. 
\\ In the absence of the electromagnetic field and of rotation, substituting the metric (\ref{metricds})
in the Einstein equations leads to a system of three
differential equations with the structure:
\be
    f' = Q_1(f,a,b,b',\Lambda) \ \ , \ \ , \ \ 
    a' = Q_2(f,a,b,b',\Lambda) \ \ , \ \ 
    b''= Q_3(f,a,b,b',\Lambda) \ \ \ \ 
\ee
the full expressions of the $Q_{1,2,3}$ are given in \cite{mrs} and
\be
        w(r) = 0 \ \ , \ \ h(r)=r^2 \ \ , \ \ g(r)=r^2
\ee 
We first discuss the solutions for $\Lambda < 0$. 
\subsection{Uniform solutions: $\Lambda < 0$}
We consider non-extremal black string solutions possessing a regular event horizon at $r=r_h$.
Near this horizon, the fields can be expanded according to   
\begin{eqnarray}
\label{eh} 
a(r)=a_h+O(r-r_h),
~
b(r)=b_1(r-r_h)+O(r-r_h)^2,~f(r)=f_1(r-r_h)+O(r-r_h)^2,{~~~~}
\end{eqnarray}
with all coefficients fixed by the parameters $a_h$, $b_1$.
Since the coordinates  $t$ and $z$ can be rescaled arbitrarily, 
the equations are invariant under arenormalizationn the functions $a(r)$ and $b(r)$.
Using this arbitrariness one can specify the four boundary
conditions for a black string according to~:
\be
f(r_h)=0 \ \ , \ \ a(r_h) = 1 \ \ , \ \ b(r_h) = 0 \ \ , \ \ b'(r_h) = 1
\ee
so that the equations can be treated as a Cauchy problem.
The profiles for $a(r)$ and $b(r)$  obtained in this way need, however, to be rescaled in such
a way that the metric is asymptotically de Sitter, Minkowski or Anti--de--Sitter (according to the value
of the cosmological constant). The asymptotic expansion of the solutions leads to 
\be
\label{asymptotic}
       f(r) = \frac{(d-1)(d-4)}{(d-2)(d-3)}+\frac{r^2}{\ell^2} + \dots \ \ , \ \
       a(r) = (\frac{d-4}{d-3})+\frac{r^2}{\ell^2}+\dots   \ \, \ \
       b(r) =(\frac{d-4}{d-3}) +\frac{r^2}{\ell^2}+\dots 
\ee
where the dots denote the various $1/r$ corrections given e.g. in \cite{br_2008}. 
\begin{figure}
\center
\includegraphics[height=6cm]{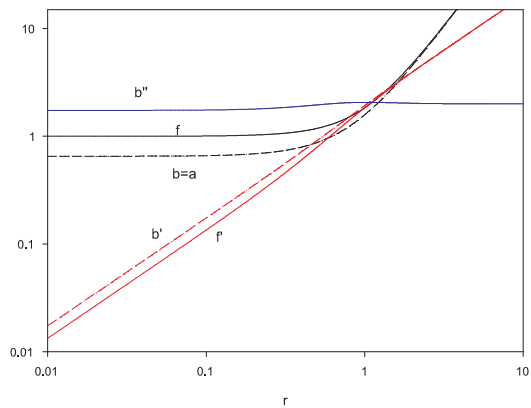}
\caption{Profiles of the regular solution for $d=8$}
\label{fig1_d6}       
\end{figure} 
\begin{figure}
\center
\includegraphics[height=6cm]{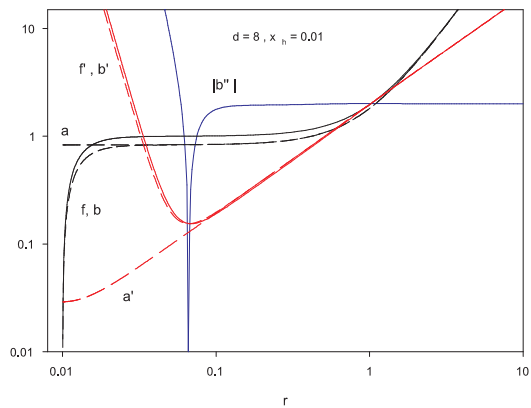}
\caption{Profiles of an AdS black string for $r_h = 0.01$}
\label{fig2_d6}       
\end{figure}  
For $\Lambda < 0$, the equations admit, to our knowledge, no explicit solutions.
Black strings were constructed numerically in \cite{mrs}. The numerical results 
strongly suggest that they  exist for arbitrary values of $r_h$. In the limit 
$r_h \to 0$, a soliton-type solution is approached.
The limiting solution has $a(r)=b(r)$ and is regular at the origin ($f(0)=1$, $f'(0)=b'(0)=0$) 
and approaches Anti--de--Sitter (AdS) space-time for $r\to \infty$. 
The convergence of the black string to the soliton is pointlike outside the origin;  
that is to say that, in the limit $r_h \to 0$, the quantities $f'(r_h),b'(r_h)$ become 
infinite  while $a(r_h)$ and $a'(r_h)$ converge 
to $1$ and $0$ respectively. 
Profiles of the regular solution and of an 
AdS black string corresponding to $r_h=0.01$ are shown in Fig.\ref{fig1_d6} and Fig.\ref{fig2_d6}, respectively.
(Note that the spike presented by  $|b'|$ in this figure is an effect of the logarithmic scale.)

\begin{figure}
\center
\includegraphics[height=6cm]{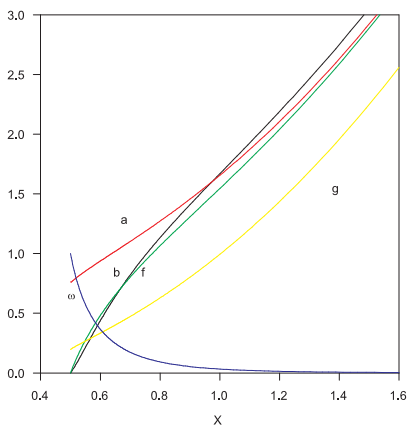}
\caption{Profile of the metric functions for a rotating AdS black string for $r_h=1$, $\Omega=1$}
\label{fig_7}       
\end{figure}

{\bf Rotating black string.} The rotating solutions have  
$w(r) > 0$ and $g(r) \neq r^2$. The results of  \cite{brs_2007} show that they exist for arbitrarily large values of $w(r_h)$. 
A typical profile of a uniform
rotating AdS black string solution is presented in Fig.\ref{fig_7}  

\begin{figure}
\center
\includegraphics[height=6cm]{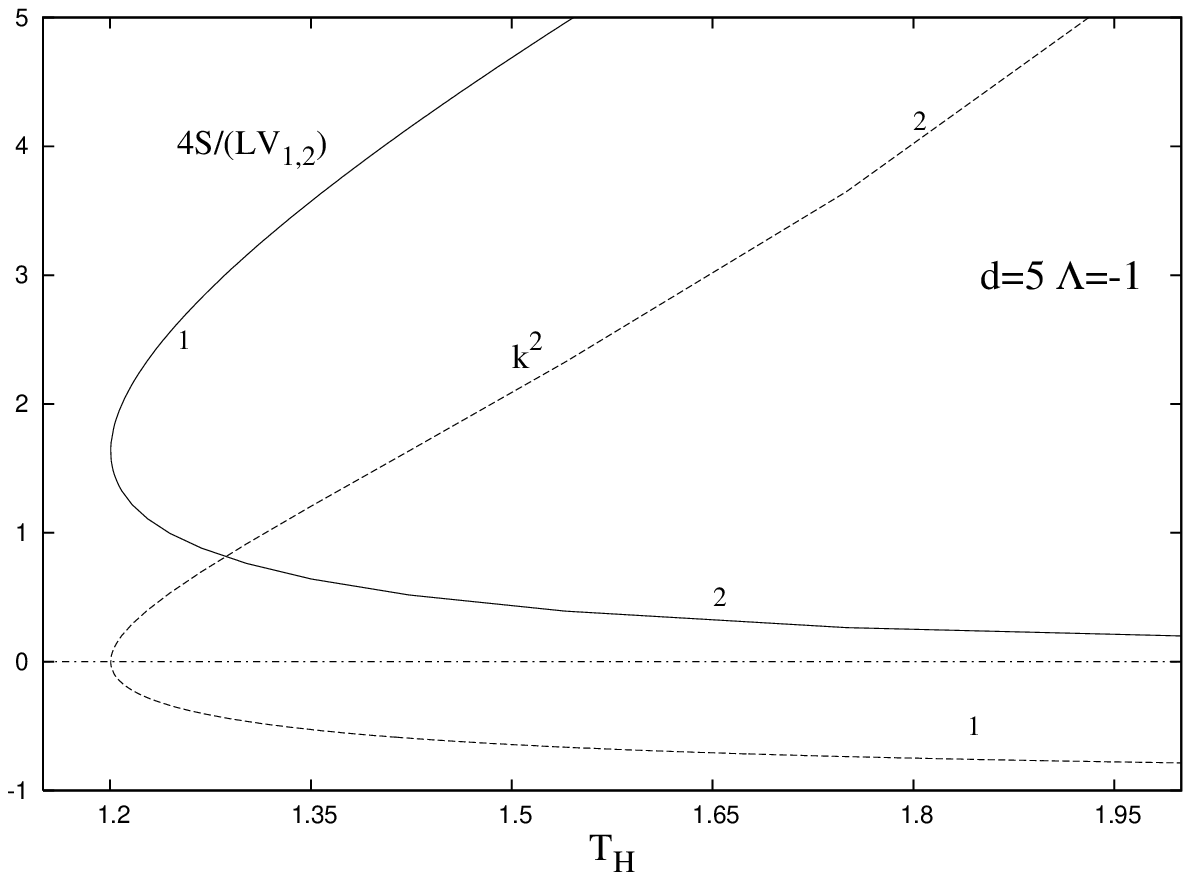}
\caption{Entropy as function of $T_H$ for the family of black strings corresponding to $\Lambda = -1$ and $d=5$}
\label{fig_8}       
\end{figure}

{\bf Physical quantities.} AdS black strings can be characterized by
conserved asymptotic charges~: their mass $M$ and tension $\cal T$. 
Using the formalism explained in detail in \cite{mrs} they can be extracted from the
asymptotic decay of the metric functions: 
\be
    M =  \frac{\ell^{d-4}}{16 \pi} L V_{d-3} [c_z - (d-2) c_t]+ M_{c}(d) 
   \ee
    \be
    a(r) = \dots + c_z (\frac{\ell}{r})^{d-3} + \dots \ ,  \ b(r) = \dots + c_t (\frac{\ell}{r})^{d-3} + \dots
\ee
Moreover, thermodynamical quantities characterizing the solutions
can also be determined.  They depend on the value of the metric at
the event horizon $r_h$~: the entropy $S$ 
\be
    S = \frac{1}{4} r_h^{d-3} L V_{d-3} \sqrt{a(r_h)}
\ee
and Hawking temperature $T_H$
\be
     T_H = \frac{1}{4} \sqrt{\frac{b'(r_h)}{r_h}((d-4) + (d-1)r_h^2/\ell^2)}
\ee
``Local thermodynamical stability'' is related to the sign of the heat capacity
\be
         C = T_H \frac{\partial S}{ \partial T_H} \ \ , \ \ {\rm for \ \ } L  {\rm \ \ fixed}
\ee  
Solutions with $C > 0$ are thermodynamically stable while those with $C<0$ are unstable. 
It is worth stating that  asymptotically flat black strings  with different $r_h$ are
related by a rescaling of the radial coordinate. On the contrary, AdS black strings with $\Lambda$ fixed
and $r_h$ varying form a family of intrinsically different solutions. 
From the analysis of \cite{mrs}, it turns out that the solutions obtained by varying $r_h$ form two
branches distinguished thermodynamically: solutions with {small $r_h$} have {$C > 0$}, 
solutions with {large $r_h$} have {$C < 0$}.
This is illustrated in Fig. \ref{fig_8} for $d=5$ and $\Lambda=-1$.

{\bf Charged black strings.} To finish this section, let us point out that charged black strings with $A = V(r) dt$ were considered
in \cite{brs_2007} as well.
The Maxwell equation
can be solved directly, leading to:
$$
    F^{tr} = \frac{q}{r^{d-3}} \sqrt{\frac{f(r)}{a(r)b(r)}} \ \ , \ \ q = {\rm constant}
$$
Charged black strings exist  for $r_h > r_{h,min} > 0$ for $q > 0$.

As main result of our analysis of the thermodynamical properties of the solutions, let us point out that 
charge and rotation change the thermodynamical stability pattern of the black strings.
For the families of solutions obtained by varying $r_h$ but fixed $L$ , $Q$
(in the case of  charged black strings ) and for fixed $L$ , $J$ (in the case of
spinning black strings, corresponding to a Grand canonical ensemble)
the unstable branch has a tendency todiminishh and disappear for large enough values of the charge
or of the angular momentum.

\begin{figure}
\center
\includegraphics[height=6cm]{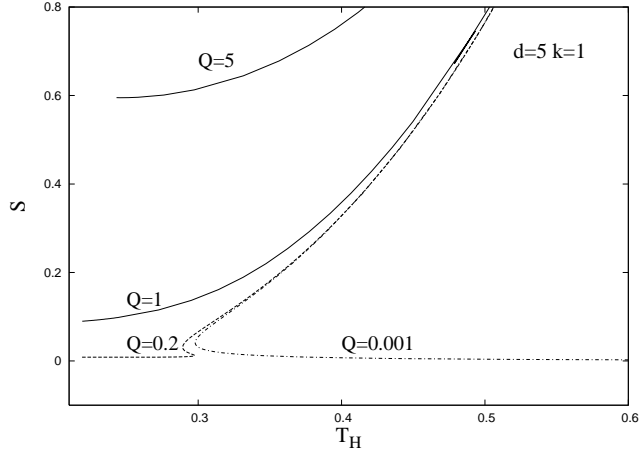}
\caption{Entropy as function of $T_H$ for families of black strings corresponding to $\Lambda = -1$ and $d=5$
with fixed electric charge }
\label{fig2bnnn}       
\end{figure} 

\begin{figure}
\center
\includegraphics[height=6cm]{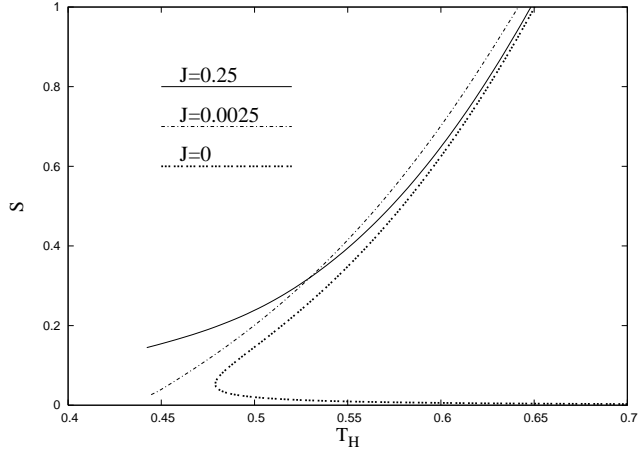}
\caption{Entropy as function of $T_H$ for families of black strings corresponding to $\Lambda = -1$ and $d=5$
with fixed angular momentum }
\label{fig2bnn}       
\end{figure} 
\subsection{\bf Uniform solutions : $\Lambda > 0$} 
\begin{figure}
\center
\includegraphics[height=6cm]{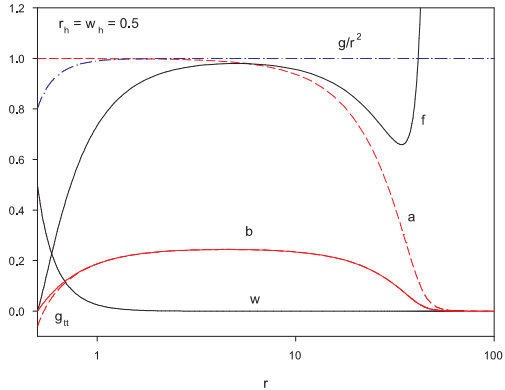}
\caption{Solution with $r_h=0.5$, $\Omega = 0.5$, $d=5$ and $\Lambda > 0$ }
\label{fig2bn}       
\end{figure}

\ \ 
Solving the equations for a positive cosmological constant \cite{bd_2007_bis}, we find
no solutions possessing both a regular horizon at $r=r_h$ and
 being asymptotically de Sitter, i.e.
\be
f(r),a(r),b(r) \to  -\Lambda r^2 + {\rm constant} + O(1/r^2) \ \ , \ \ {\rm for} \ \ \ r \to \infty 
\ee
\\
As pointed out above, imposing a  regular horizon at $r = r_h$ needs only conditions 
at the horizon. Extrapolating the initial data up to $r \to \infty$
reveals that the solution  evolves asymptotically  into a configuration such that
\be
\label{kasner}
a(r) \to r^{\alpha} \ ,  \ b(r) \to r^{\alpha} \ , \ f(r) \to r^{\gamma} 
\ee
where the parameters $\alpha$, $\gamma$ depend on $d$~: 
\be
     \alpha =  -2(d-3) - \sqrt{2(d-2)(d-3)}, \gamma = 2(d-2) + \sqrt{2(d-2)(d-3)} \ \ . 
\ee
The form (\ref{kasner}) 
corresponds  to one possible asymptotic behaviour of the solutions, the other possibility is de Sitter.
A solution for $r_h=0.5$ (including rotation) is presented in Fig. \ref{fig2bn}.
Examining the Kretschmann scalar reveals that the solution 
with the asymptotics (\ref{kasner}) are singular at $r = \infty$.
Along with the case $\Lambda < 0$, solutions regular at the origin exist as well, but have the asymptotics (\ref{kasner}).
To finish this discussion, we mention that imposing a regular cosmological horizon at $r = r_c$ leads to the absence 
of a regular horizon at finite distance and to a naked singularity at the origin.

Black strings with $\Lambda \neq 0$ therefore lead to a situation where no analytical continuation of the $\Lambda < 0$ solutions into the $\Lambda >0$ domain can be established;  
it is tempting to relate this result to the absence of explicit solutions.
\subsection{Stability of AdS black strings}
\begin{figure}
\center
\includegraphics[height=6cm]{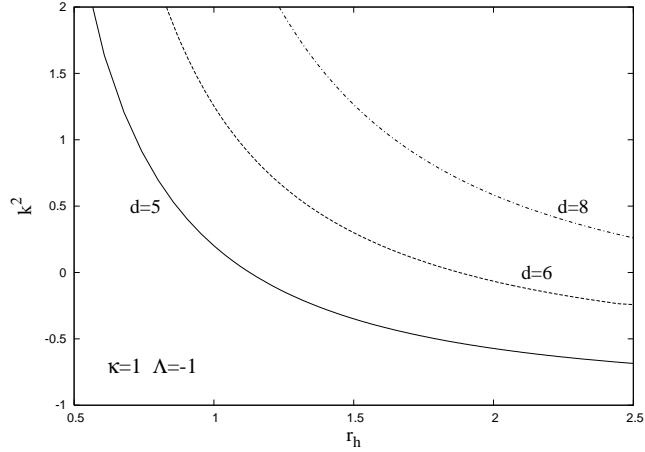}
\caption{The value of $k^2$ as a function of $r_h$ for $d=5,6,8$ and $\Lambda = -1$}
\label{fig_sta1}       
\end{figure}

\begin{figure}
\center
\includegraphics[height=6cm]{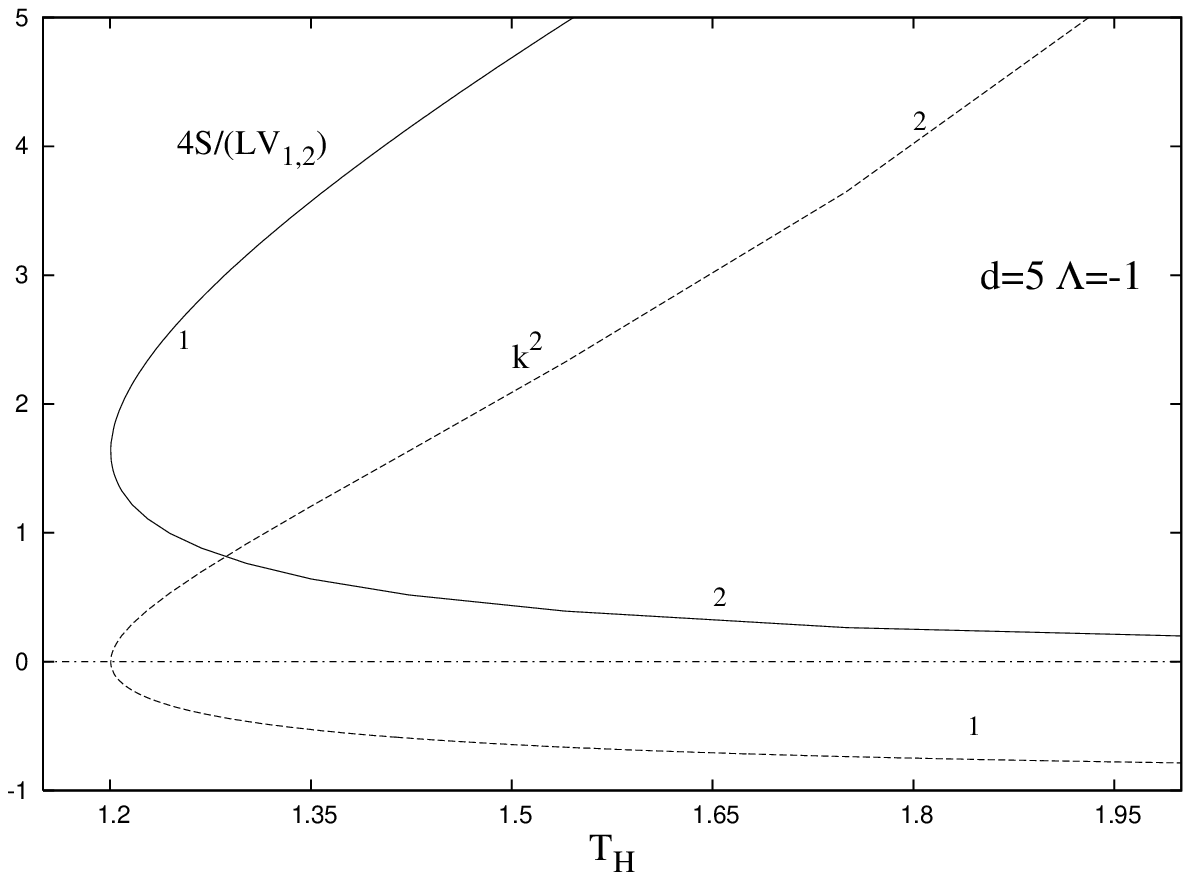}
\caption{The entropy $S$ and the eigenvalue $k^2$ as function of $T_H$ for $d=5$. The figure
shows that the GM conjecture is obeyed. }
\label{fig_sta2}       
\end{figure}

\begin{figure}
\center
\includegraphics[height=6cm]{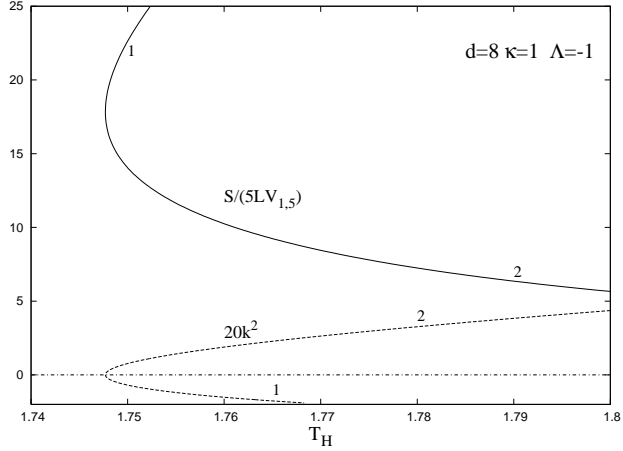}
\caption{The entropy $S$ and the eigenvalue $k^2$ as function of $T_H$ for $d=8$. The figure
shows that the GM conjecture is obeyed. }
\label{fig_sta3}       
\end{figure}

As mentioned above, the extra dimension of space-timeparameterizedd by $z$ 
is assumed to be periodic with period $L$,  i.e.  $z \in [0,L]$.
One of the most striking facts about  asymptotically flat black strings  
is that they present an instability \cite{gl} for large values of $L$, as discovered by Gregory and Laflamme (GL).
It is therefore natural to determine whether AdS black strings also have  an instability of the GL 
type and to attempt to relate this eventual instability to the thermodynamical instability discussed in 
previous sections. In this framework, it is interesting to check whether the following conjecture 
formulated by Gubser and Mitra
\cite{gm} is fulfilled for the ADS black strings~:
{\it  "For a black brane solution to be free of dynamical instabilities
it is necessary and sufficient for it to be locally thermodynamically stable."}
 \\
{\bf Digression to the electroweak model.}
To some extend theoccurrencee of instabilities of uniform black strings and the emergence of
non--uniform (i.e. $z$-depending) solutions can be compared with an older problem: the sphaleron instability.
The classical equations of the SU(2)$\times$ U(1) Yang-Mills-Higgs (the bosonic sector of the electroweak Lagrangian)
 admit a solution: the Klinkhamer-Manton (KM) sphaleron \cite{km_1984}. The study of the instability of the
 sphaleron can be performed by linearizing the equations about the sphaleron with respect to a time-dependent
 fluctuation, i.e. 
\be
      \Phi(t,r) = \Phi_{KM}(r) + e^{\omega t} \eta(r) \longrightarrow H \eta = \omega \eta
\ee
where $\Phi_{KM}$ symbolizes the sphaleron configuration. 
All physical parameters can be scaled away from the equations apart from the Higgs-particle mass $M_H$
which is left as a free parameter (we assume the limit of vanishing Weinberg angle $\theta_w = 0$).
It was shown \cite{bk_1988,yaffe_1988,bk_1990} that the linearized equations lead to a spectral problem with spectral
parameter $\omega$ and that normalizable fluctuations exist for  specific eigenvalues $\omega(M_H)$.
For a series of critical values  $M_{H,c}$, we have zero modes  $\omega(M_{H,c})=0$ 
and the number of {negative} eigenvalues depends on $M_H$. 
As a consequence new solutions exist for $M_H \geq M_{H,c}$ bifurcating from the KM sphaleron.
 Because these new solutions appear in pairs, related to each other by parity, we called
 them {\it bisphaleron}.
The analogies between the 20--year old bisphaleron and the more recent non--uniform black strings
can be seen as follows~:
\begin{itemize}
\item Yang-Mills-Higgs equations {$\longrightarrow$}  Einstein equations for $d > 4$.
\item Higgs field mass $M_H$ {$\longrightarrow$} Length of the co-dimension $L$.
\item KM-sphaleron {$\longrightarrow$} Uniform Black String (Schwarzschild).
\item bisphaleron {$\longrightarrow$} non--uniform black string.
\item Breaking of parity {$\longrightarrow$} Breaking of the translation symmetry in $z$. 
\item Morse theorem + Catastrophe theory {$\longrightarrow$} Gubser-Mitra conjecture.
\end{itemize}
Coming back to the GL-instability for AdS black strings, we consider a deformation of the metric
which we parametrize like in \cite{Gubser:2001ac}:
\be
ds^2 = -b(r)e^{2A(r,z)}dt^2 + e^{2B(r,z)}\left(\frac{dr^2}{f(r)} + a(r)dz^2\right) 
  + r^2e^{2F(r,z)}d\Omega_{d-3}^2 \ \ ,
\ee
assuming no time-dependence of the metric fields. That is to say that we specialize
in the zero-mode, i.e. $\Omega=0$ where $\Omega$ is used in \cite{gl}. The functions $A,B,C$ 
encode the deviations with respect to the uniform metric and depend on $z$ and $r$. Using the
periodic conditions assumed in $z$, the fluctuations can be expanded in Fourier series~:
\begin{eqnarray*}
X(r,z) = \epsilon X_1(r)\cos(k z) + \epsilon^2 (X_0(r) + X_2(r) \cos(2 k z) ) + O(\epsilon^2) \ , 
\ k \equiv \frac{2\pi}{L}
\end{eqnarray*}
where $X$ stands for  $A,B,F$    
 and $\epsilon$ denotes an infinitesimal parameter.
 Extracting the  terms linear in
$\epsilon$ from the Einstein equations  leads to a system of linear differential equations in $A_1(r),B_1(r),F_1(r)$. 
Instead of writing these lengthly equations (they can be found in  \cite{bdr_2007}), we point out a few of
their properties~:
\begin{itemize}
\item The ``potentials'': the functions $a(r),b(r),f(r)$ are known only numerically.
\item The perturbations $X_1(r)$ should vanish asymptotically.
\item The function $B_1(r)$ can be eliminated from the system.
\item Regularity at $r=r_h$ leads to two conditions of the form $\Gamma(A,C,A',C')(r=r_h) = 0$.
\item  Special values of  $k^2$ have to be determined such that boundary conditions are fulfilled
up to a global factor~: it is an eigenvalue problem.
\item A rescaling of the radial coordinate can be used to  set either $r_h$ or $\Lambda$
to a canonical value, e.g. $\Lambda = -1$.
\item  Solutions with $k^2 > 0$ are unstable, $k^2 < 0$ are stable.
\end{itemize}
Fig. \ref{fig_sta1} summarizes our results for the critical value of $k$ as a function of the event horizon value
$r_h$ for several values of $d$.
The comparison between the entropy of the solution and the eigenvalue $k^2$ as functions of the
Hawking temperature $T_H$ is reported in Figs.\ref{fig_sta2} and \ref{fig_sta3} for $d=5,8$ respectively.
These figures clearly show that the solutions having $k^2 > 0$  have
a negative heat capacity and are therefore unstable, while solutions with $k^2 <0$
have a positive heat capacity and are stable.
The fact that the change of sign of $k^2$ coincides with the change of sign of
the heat capacity indicates that the GM conjecture is fulfilled for AdS black strings.
\subsection{Non-uniform (EGB) AdS black strings.}
The results of the previous section strongly suggest that non-uniform black strings
with AdS asymptotics should exist as well. To construct such solutions, the above spherically symmetric
ansatz has to be generalized in order to allow the metric to depend on the radial variable $r$ and
on the extra dimension denoted here by $y$.

For the calculations, it is convenient to use an alternative parametrisation of the radial variable 
and to work with a new one, $\tilde r$, defined through
\be
ds^2 = - e^{2 A} \tilde b dt^2 + \tilde g(\tilde r) e^{2 C} d \Omega ^2 
+ e^{2 B} (\frac{1}{\tilde f} \frac{\tilde r^2}{\tilde g} dr^2 + \tilde a dy^2) 
\ee
where the "tilde" refers to quantities depending on the radius $\tilde r$. The dependence on $y$ appears
through the functions $A,B,C$.
Practically, we use $\tilde g = \tilde r^2 + r_h^2$. The event horizon thencorrespondss to $\tilde r = 0$
and the relations   $\tilde f(\tilde r) = f(r)$,$\tilde a(\tilde r)=a(r)$,$\tilde b(\tilde r)=b(r)$ hold.

 The system of partial differential equations for $A,B,C$ has to be solved with 
 the following boundary conditions 
  \be
                  \partial_r A(0,y)= \partial_r C(0,y)=0 \ \ , \ \ B(0,y)-A(0,y)= \alpha \ \ , \ \ \ X(\infty,y)=0
 \ee
 \be
                  \partial_y X(r,0)=0 \ \ , \ \ \partial_y X(r,L)=0 \ \ , \ \ X = A , B , C
 \ee
 where the parameter $\alpha$ will enforce the deformation with respect to uniform configurations. 
 The solutions that we are looking for are periodic for $y \in [0,L]$ and present a mirror symmetry  $y \to L-y$.
 The corresponding  equation was solved numerically in the case $\Lambda=-1$, $r_h=1$ 
 corresponding to $k_{cr}\approx 1.143$
 (determining the critical radius $L_{cr}=2\pi/k_{cr}$) and for several values of $\alpha$. 
 For the numerical integration, we used a compactified variable $x = \tilde r / (1 + \tilde r)$ and $z=y/L_{cr}$.
 The square $(x,z) \in [0,1] \times [0,1]$ was discretized with grids of 80 $\times$ 80 points.
 
\begin{figure}
\center
\includegraphics[height=6cm]{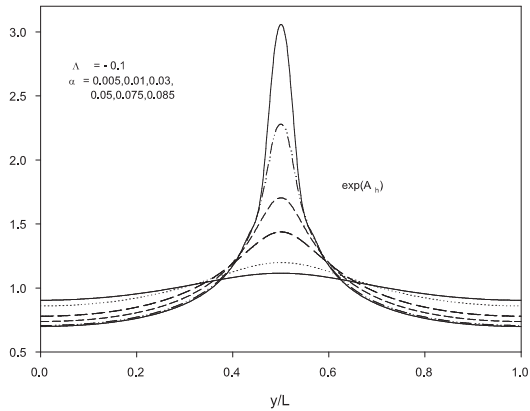}
\caption{The profiles of the quantity $\exp A$ evaluated at the horizon for $\Lambda = - 0.1$  
and several values of $\alpha$ and $d=6$}
\label{nubs_1}       
\end{figure}

\begin{figure}
\center
\includegraphics[height=6cm]{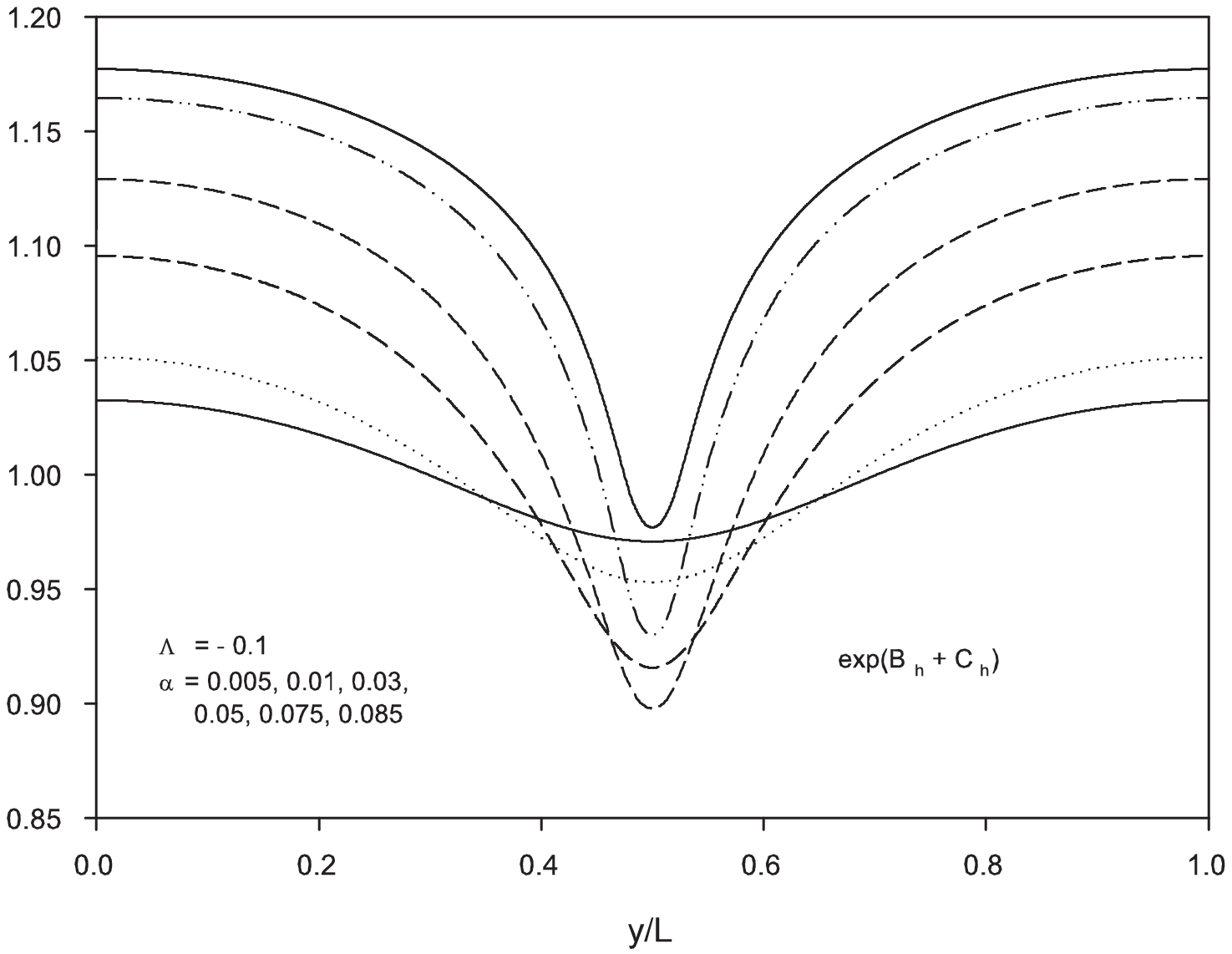}
\caption{The profiles of the quantity $\exp B+C$ evaluated at the horizon for $\Lambda = - 0.1$  
and several values of $\alpha$ and $d=6$}
\label{nubs_2}       
\end{figure}

\begin{figure}
\center
\includegraphics[height=6cm]{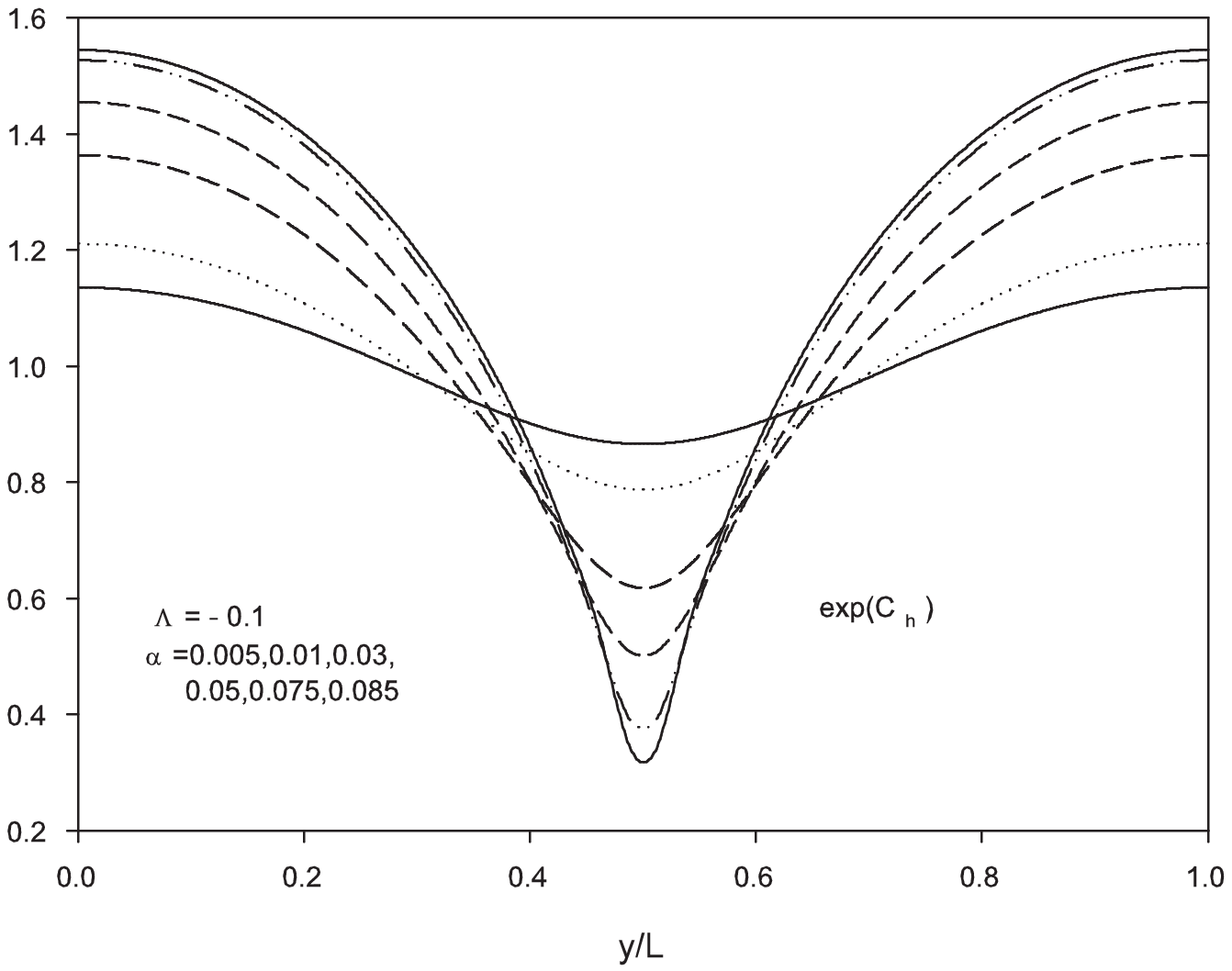}
\caption{The profiles of the quantity $\exp C$ evaluated at the horizon for $\Lambda = - 0.1$  
and several values of $\alpha$ and $d=6$}
\label{nubs_3}       
\end{figure}

\begin{figure}
\center
\includegraphics[height=6cm]{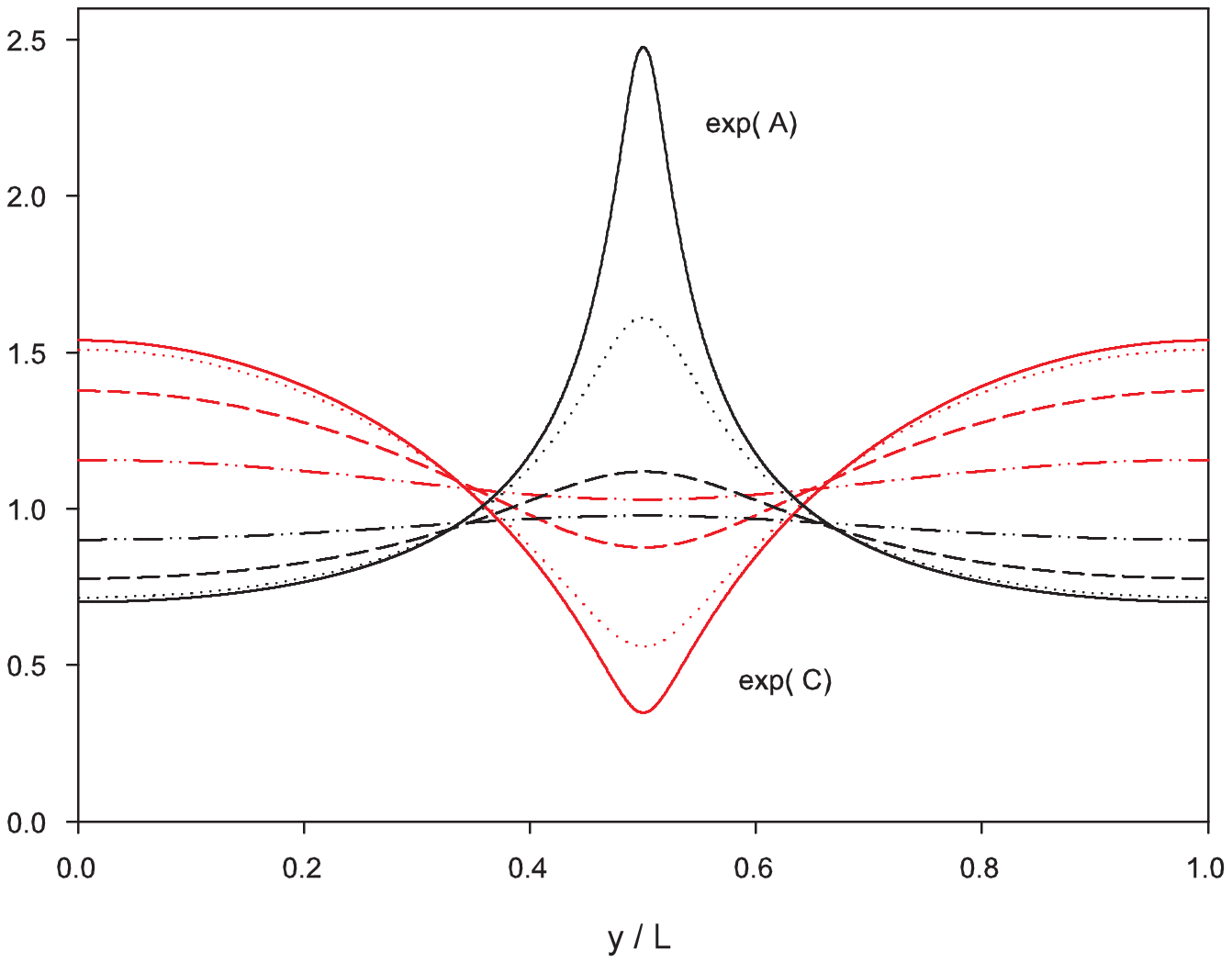}
\caption{The profiles of the values $\exp A(x,y)$ (black lines)  and $\exp C(x,y)$ (red lines)
 as functions of $y$ and for $x=0,0.2,0.4,0.6$ and $\Lambda = - 0.1$  }
\label{nubs_08_ac}       
\end{figure}

\begin{figure}
\center
\includegraphics[height=6cm]{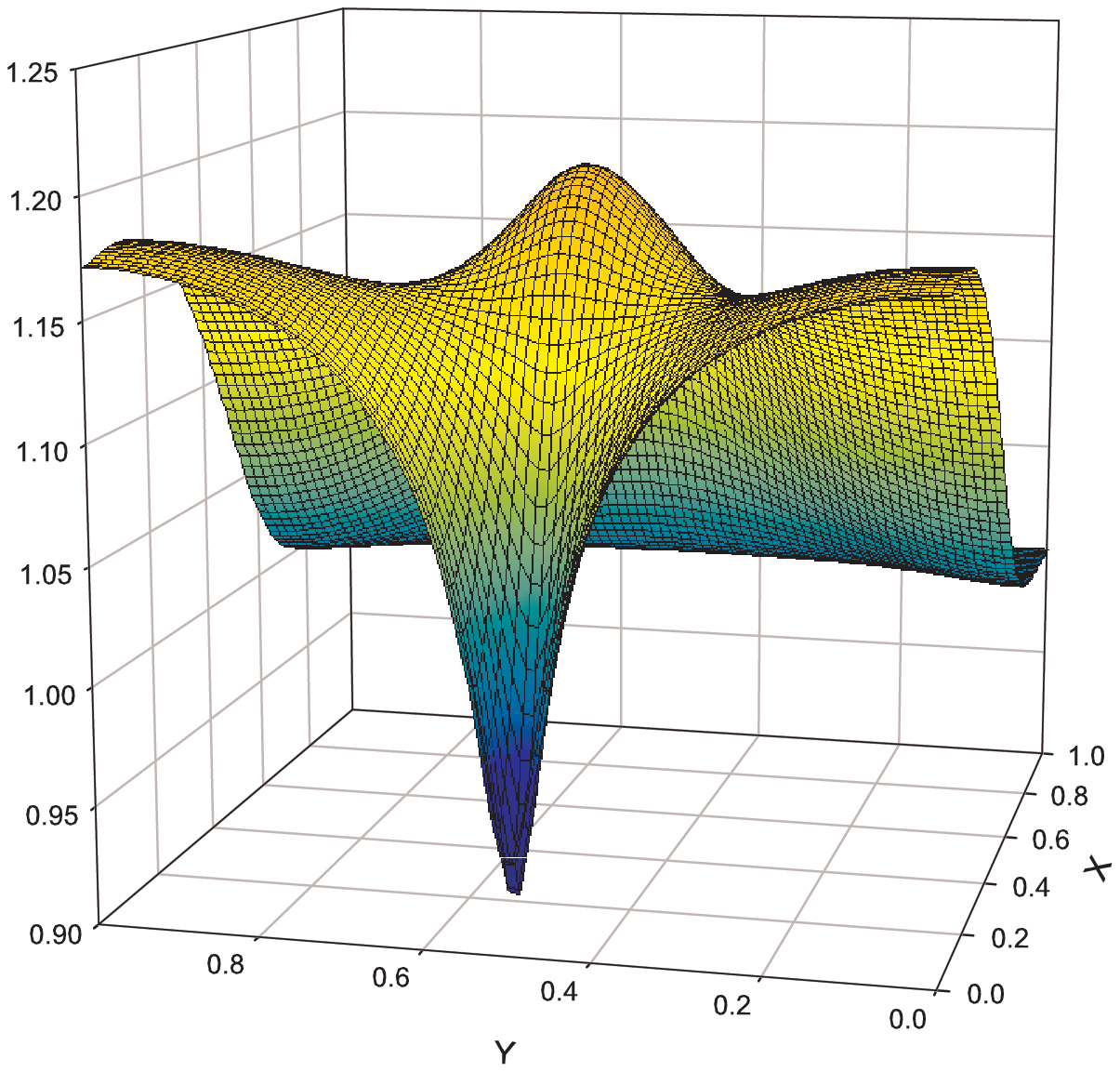}
\caption{Two-dimensional plot  of the function $\exp (B+C)(x,y)$ for $\alpha=0.8$ and $\Lambda = - 0.1$}
\label{nubs_08_bc}       
\end{figure}

 Profiles of the deformation functions $A,B,C$  are given in several figures.
 In  Fig.\ref{nubs_1} the deviation of the $g_{tt}$ metric component at the horizon is given as function of $y$ for
 several values of $\alpha$. Similar plots of the quantity $\exp(B+C)$  are given in Fig.\ref{nubs_2}. 
 This quantity is directly relevant for  the calculation of the entropy of the non-uniform black strings. 
 Finally, the quantity $\exp(C)$, encoding  the
  deformation parameter introduced e.g. in \cite{Gubser:2001ac} is given in Fig. \ref{nubs_3}.
  The figures reveal that, for $\alpha$ increasing, the deformation becomes 
 very pronounced at $y=L/2$. The grids used where not sufficient to get reliable solutions for $\alpha > 0.9$.
 More detailed studies of these non-uniform solutions are presented in \cite{delsate_2008}.
The dependence of the solutions on the radial variable  is further illustrated 
in Figs.\ref{nubs_08_ac} and  \ref{nubs_08_bc} correspondingg to $\alpha = 0.8$)
The functions $A$ and $C$ decrease monotonically to zero for $r\to \infty$. The behaviour
of the combination $B+C$ is more involved since it presents a local maximum  
 at an intermediate value $r=r_m$ for $y=L/2$ . 
\section{AdS black strings in Einstein-Gauss-Bonnet}
So far, we presented blackobjectssoccurringg in minimal models for gravity, i.e. constructed within
the minimal Einstein-Hilbert action. In higher dimensions, however, there exist more general choices of 
physically acceptable Lagrangians describing gravity. Low energy effective models descending
from string theory contain such terms. It is therefore natural to pay attention to the influence of
additional terms in the gravity sector on the classical solutions i.e. on black holes and black strings.
The Gauss-Bonnet (GB) interaction is the first curvature correction to General Relativity from
the low energy effective action of string theory.
It is  quartic  in the metric and of second order in the derivatives. 
\subsection{The Einstein-Gauss-Bonnet equations}
In this section, 
we consider the Einstein-Gauss-Bonnet (EGB) action supplemented
with a  cosmological constant $\Lambda=-(d-2)(d-1)/2\ell^2$~: 
\begin{eqnarray}
\label{action}
I=\frac{1}{16 \pi G}\int_\mathcal{M}~d^dx \sqrt{-g} \left(R-2 \Lambda+\frac{\alpha}{4}L_{GB} \right),
\nonumber
\end{eqnarray}
$R$ is the Ricci scalar and 
\be
\label{LGB}
L_{GB} =R^2-4R_{\mu \nu}R^{\mu \nu}+R_{\mu \nu \sigma \tau}R^{\mu \nu \sigma \tau}, \nonumber
\ee
denotes the Gauss-Bonnet term.
Variation of this action with respect to the metric results in the EGB equations: 
\begin{eqnarray*}
\label{eqs}
R_{\mu \nu } -\frac{1}{2}Rg_{\mu \nu}+\Lambda g_{\mu \nu }+\frac{\alpha}{4}H_{\mu \nu}=0~, \nonumber
\end{eqnarray*}
where
{\begin{equation}
\label{Hmn}
H_{\mu \nu}=2(R_{\mu \sigma \kappa \tau }R_{\nu }^{\phantom{\nu}%
\sigma \kappa \tau }-2R_{\mu \rho \nu \sigma }R^{\rho \sigma }-2R_{\mu
\sigma }R_{\phantom{\sigma}\nu }^{\sigma }+RR_{\mu \nu })-\frac{1}{2}%
L_{GB}g_{\mu \nu }  ~. \nonumber
\end{equation}}

It is useful to define an effective Anti-de-Sitter radius by means of
{\begin{eqnarray}
\label{lc}
\ell_c=\ell\sqrt{\frac{1+U}{2}},~~{\rm~~with~~~~}U=\sqrt{1-\frac{\alpha(d-3)(d-4)}{\ell^2}},
\nonumber
\end{eqnarray}}
\\
This combination of $\ell$ and $\alpha$ indeed appears naturally in the
asymptotic expansion of the solution and in  the counterterm  formalism discussed in the next section.
The occurrence of $\ell_c$ in the equation has consequences on the solutions since it
leads to the existence of an upper bound for the  Gauss-Bonnet coefficient,
$\alpha\leq \alpha_{max}=\ell^2/(d-3)(d-4)$ in the case of asymptotically AdS solutions.

\subsection{Counterterm Formalism}
Several times in the previous sections of this manuscript we used tacitely the existence of regularizing counterterms
which allow the theory to be well defined and finite. This is not misleading since these countertems
do not affect the classical equations. Here,  we will present their explicit form and illustrate their interest
for the theory under investigation, i.e. the Einstein-Gauss-Bonnet model.
   
Let us first remark that the various solutions of the models considered here do not 
have a finite action because of the non-compact character of space-times with $\Lambda = 0$ or $\Lambda < 0$. 
In order to enforce  finite numbers for the action, one technique
consists in adding suitable counterterms to the original action \cite{Balasubramanian:1999re,Brown:1993}.
The counterterms are constructed in such a way that the full Lagrangians fulfill several requirements, namely: 
\begin{itemize}
\item  They depend on curvature invariants associated with the geometry at the boundary of space-time.
\item They do not affect the equations.
\item They are also infinite, in order to cancel the divergences of the basic action.
\end{itemize}
As a byproduct, the counterterms lead to a boundary stress tensor $T_a^b$ 
which allows in particular to define conserved quantities like mass and angular momentum \cite{ghezelbash_mann}.\\
The counterterms are known in the case of the  Einstein-Hilbert action; 
we present here the generalization of this result to the case of  the Einstein-Gauss-Bonnet action.
  

For $d<8$ and even, the appropriate counterterms are given by  \cite{br_2008}
\begin{eqnarray*}
\label{Lagrangianct} 
I_{\mathrm{ct}}^0 &=&\frac{1}{8\pi G}\int_{\partial \mathcal{M}} d^{d-1}x\sqrt{-\gamma 
} \\ \nonumber
\bigg\{
  &-&(\frac{d-2}{\ell_c })(\frac{2+U}{3})
 -\frac{\ell_c \mathsf{\Theta } \left( d-4  \right) } {2(d-3)}(2-U) {\mathsf{R}}
 \\
 \nonumber
 &&-\frac{\ell_c ^{3}\mathsf{\Theta }\left( d-6\right) }{2(d-3)^{2}(d-5)} \\ \nonumber
 &&\left[ U\bigg( {\mathsf{R}_{ab}\mathsf{R}^{ab}}-\frac{d-1}{4(d-2)} {\mathsf{R}^{2}} \bigg)
 -\frac{d-3}{2(d-4)}(U-1) {L_{GB}} 
  \right] 
\bigg\},
\end{eqnarray*}
where the quantity $U$ is defined above and
\begin{itemize}
\item $\gamma$ in the induced metric of the boundary of space-time.
\item {$\mathsf{R}$, $\mathsf{R}^{ab}$}  and {$L_{GB}$} are the curvature, the 
Ricci tensor and the Gauss-Bonnet term associated with $\gamma $. 
\item  $\mathsf{\Theta}(x)$ is the step-function with  $\mathsf{\Theta 
}\left( x\right) =1$ provided $x\geq 0$, and zero otherwise. 
\end{itemize}
Up to the cases we have addressed, it turs out that these counterterms 
appears as a truncated series of powers of $R_{abcd}$ and $\ell_c$. We guess this can be
generalized to arbitrary values of $d$ although we have no formal proof of this. 
Let us finally stress that the corresponding counterterms of Einstein gravity are 
recovered for $\alpha \to 0$ (i.e. $U \to 1$) and that  
for odd values of $d$ the expression is  more involved. 
\\
\subsection{Black string and thermodynamical properties}
\begin{figure}
\center
\includegraphics[height=6cm]{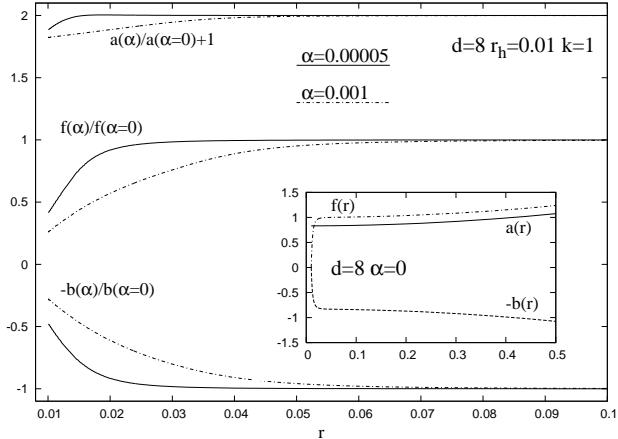}
\caption{Comparison of profiles of E-BS and EGB-BS for  
small values of $\alpha$ and $d=8$}
\label{fig_gb1}       
\end{figure}

\begin{figure}
\center
\includegraphics[height=6cm]{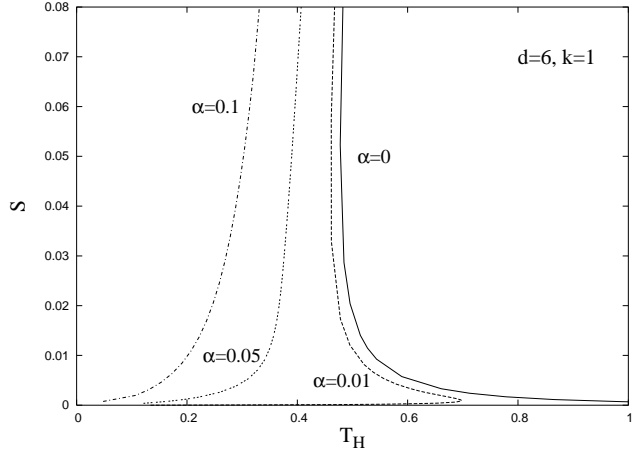}
\caption{The entropy $S$ as a function of $T_H$ for  different
values of $\alpha$ and $d=6$}
\label{fig_gb2}       
\end{figure}

We now discuss the black string solutions of the Einstein-Gauss-Bonnet (EGB) equations.
At first sight, these solutions appear
as smooth deformations of the Einstein black strings, although the deviation from the pure Einstein black strings
is systematically significant even for infinitesimal values of $\alpha$. This is seen in Fig. \ref{fig_gb1}.
One striking property is that the Smarr relation available for $\alpha = 0$ is obeyed for $\alpha >0$
\be
    M + {\cal T} L = T_H S
\ee
The  effect of the Gauss-Bonnet (GB) terms appears more drastically when looking at  
the curve $S(T_H)$,  (shown in Fig. \ref{fig_gb2}) demonstrating  the influence 
of the GB  interaction on the thermodynamical properties: the unstable  branch of solutions
occurring for $\alpha = 0$ disappears progressively in favour of a branch of thermodynamically 
stable solutions when the Gauss-Bonnet coupling constant $\alpha$ increases.
\\
\subsection{Domain of existence}
In this section, we discuss the domain of existence of the EGB black strings
in terms of the parameters $\alpha, r_h$ and $\Lambda$.
In the case $\Lambda = 0$, the black strings solutions are discussed in \cite{kobayashi_tanaka:2005} for $d=5$.
Besides black strings,   $M_{d-1}\times S_1$  constitutes a regular solution of the equations on $r\in [0,\infty]$ 
irrespectively of the value of $\alpha$ ($M_p$ denotes $p$-dimensional Minkowski space). 

As mentioned above, for $\Lambda \neq 0$,
the black strings of the vacuum Einstein equations exist for arbitrary values
of $r_h$ and approach soliton-type solution in the limit $r_h \to 0$.
The limiting solution has $a(r)=b(r)$, it is regular at the origin ($f(0)=1$, $f'(0)=b'(0)=0$) 
and approaches Anti--de--Sitter space-time for $r\to \infty$. 
The convergence of the black string to the soliton  is therefore pointlike outside the origin;  
that is to say that, for $r_h \to 0$, the quantities $f'(r_h),b'(r_h)$ become 
infinite  while $a(r_h)$ and $a'(r_h)$ converge  to $1$ and $0$ respectively. The corresponding
curves are shown (for $d=6$) (red lines) in Fig. \ref{rh_vary}.
\\ Determining the domain of existence of black strings in the  EGB case
 needs a detailed analysis of the behaviour of solutions in the limit $r_h \to 0$.
EGB black strings were studied in \cite{br_2008} but details about their behaviour
in the  $r_h\to 0$ limit will be reported here. As we will see, the pattern 
crucially depends on  the number of dimensions.
\subsubsection{Case d=5}
In this case, the solutions exist only on a sub-domain of the $r_h$-$\alpha$ plane limited by
 $\alpha < \alpha_m$ with  $\alpha_m = r_h^2/ (2(1+r_h^2))$. 
Performing the expansion (\ref{eh}) about the event horizon, 
leads for  the parameter  $f_1$ to 
 \begin{eqnarray}
\label{d5f1} 
f_1=\frac{r_h^2(\ell^2+2\alpha)}{\ell^2 \alpha}
-\frac{1}{\ell^2 \alpha}\sqrt{(\ell^2-2\alpha)\left(r_h^2(\ell^2-2\alpha)-2\ell^2\alpha \right)},
\end{eqnarray}
which clearly implies that real solutions exist for $r_h > \ell/\sqrt{\ell^2/(2\alpha)-1}$.
\subsubsection{Case d=6}
\begin{figure}
\center
\includegraphics[height=6cm]{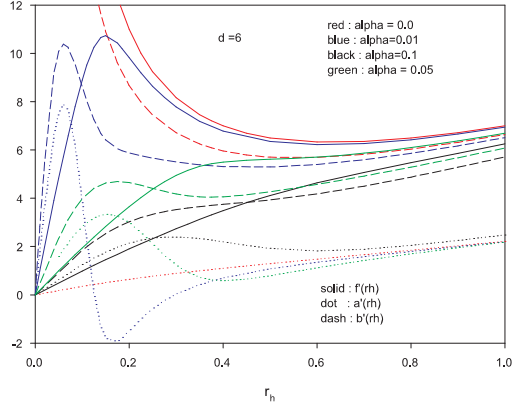}
\caption{The parameters at the horizon of some d=6 solutions as functions of $r_h$ for two different
values of $\alpha$}
\label{rh_vary}       
\end{figure}

\begin{figure}
\center
\includegraphics[height=6cm]{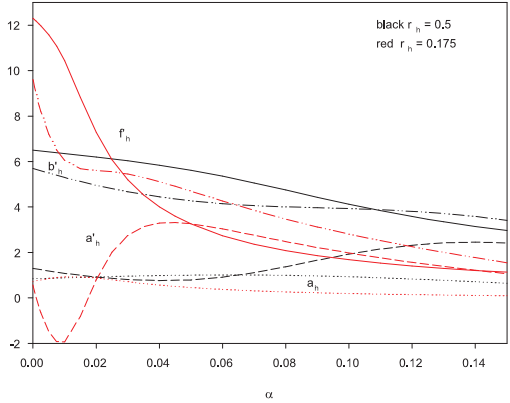}
\caption{Some parameters at the horizon of some d=6 solutions as functions of $\alpha$ for two different
values of $r_h$: $f'$ (solid), $a$ (dots), $a'$ (dash), $b'$ (dot-dash)}
\label{alpha_vary}       
\end{figure}

The expression (\ref{d5f1}) of $f_1$ for generic $d$ is much more complicated and does not bring definite information
about the domain for $d>5$.
The domain of existence of 6-dimensional EGB black string is more tricky, as illustrated by  
Figs. \ref{rh_vary} and \ref{alpha_vary}.
First, let us mention that the solutions exist only for  $\alpha < 1/6$ which constitute the main critical value.
To understand the pattern of the solutions, it is worth looking at the parameters $f'_h,a'_h,b'_h$.
For a fixed positive value of $\alpha$  the quantities $a'_h,b'_h,f'_h$ behave like in the $\alpha = 0$ case (red lines)
for large $r_h$. When $r_h$ diminishes, they deviate from their  values in the Einstein case, they attain a maximum and then
all decrease to  zero for $r_h \to 0$. This suggests that the  EGB black strings approach a configuration with a singularity 
at the origin in the limit $r_h \to 0$. 
Figure (\ref{alpha_vary}) further illustrates how the parameters $f'(r_h),a(r_h),a'(r_h),b'(r_h)$
 vary as functions of $\alpha$ for two different values of the horizon.
For large horizon values, e.g.  $r_h \sim 0.5$, these variations are small. For smaller $r_h$  
the variations are more significant and some oscillations are observed.
These results reveal a  non-perturbative character of the Gauss-Bonnet coupling constant~:
a small variation of  $\alpha$ leads to a significant change in the profiles of the metric functions
and especially of their derivatives. 
The correction is likely non polynomial and cannot be treated perturbatively.
\subsubsection{Case d=8}
\begin{figure}
\center
\includegraphics[height=6cm]{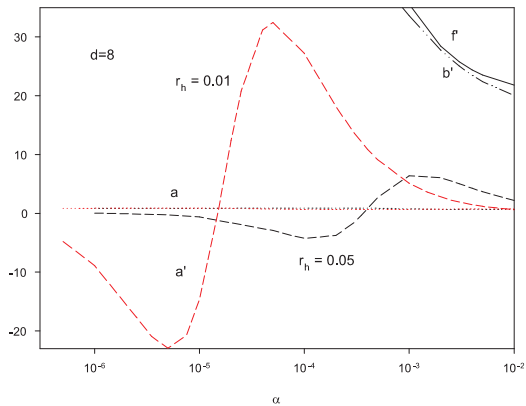}
\caption{Values of $a_h$ and $a'_h$ as functions of $\alpha$ for different values of $r_h$}
\label{aprime_d8}       
\end{figure}

The analysis of the $r_h \to 0$ limit in the case $d=8$ is even more subtle and undefinite. 
Our numerical results do not confirm the existence of a definite solution in this limit 
(even presenting  a singularity at the origin).
It should be stressed that it turns out to be extremely  difficult to construct numerical solutions for $\alpha >0$
and $r_h < 0.01$. 
Examining  the behaviour of the derivatives of $f,b$ at the horizon  
(say $f'_h$ and $b'_h$) for fixed positive $\alpha$ and varying $r_h$ 
reveals that these quantities increase when $r_h$ decreases,
 like for Einstein-black strings in $d=6$ (see Fig. \ref{rh_vary}). 
The situation is, however, different because the value $a(r_h)$  
 (see  Fig. \ref{aprime_d8})  varies  
non monotonically. It develops  some oscillations
 when both $\alpha$ and $r_h$ are small.
 Our results suggest that these oscillations become more and more pronounced
when the value of $r_h$  decreases. This makes it not easy to construct these solutions 
numerically and it turns out impossible to have proper insight into the nature of the  
limiting configuration.

\section{Conclusions}
 We have constructed black holes and black strings solutions within several models
in the presence of a cosmological constant. 
 Up to our knowledge, the extensions that we have 
discussed do not allow explicit solutions of the equations. 
 We therefore used numerical
methods to solve the equations. 
 We hope these results contribute to a more general understanding
of the classification of solutions of the Einstein equations in $d > 4$. 
\\
\\
{\bf Acknowledgments.} It is a pleasure to acknowledge J. Kunz and C. L\"ammerzahl for their invitation to the
Heraeus Seminar in Bremen in August 2008. I am also grateful to T. Delsate
and E. Radu for their collaboration on the topic. 
\\
\\

\end{document}